\documentstyle[12pt]{article}
\def\hybrid{\topmargin 0pt \oddsidemargin 0pt \headheight 0pt \headsep 0pt
	\textheight 9in 
        \textwidth 6.25in 
	\marginparwidth .875in \parskip 5pt plus 1pt \jot = 1.5ex}

 \catcode`\@=11 \def\marginnote#1{}
\newcount\hour \newcount\minute \newtoks\amorpm \hour=\time\divide\hour
by60 \minute=\time{\multiply\hour by60 \global\advance\minute by-\hour}
\edef\standardtime{{\ifnum\hour<12 \global\amorpm={am}%
\else\global\amorpm={pm}\advance\hour by-12 \fi \ifnum\hour=0 \hour=12 \fi
\number\hour:\ifnum\minute<10 0\fi\number\minute\the\amorpm}}
\edef\militarytime{\number\hour:\ifnum\minute<10 0\fi\number\minute}

\def\draftlabel#1{{\@bsphack\if@filesw {\let\thepage\relax
   \xdef\@gtempa{\write\@auxout{\string
   \newlabel{#1}{{\@currentlabel}{\thepage}}}}}\@gtempa \if@nobreak
   \ifvmode\nobreak\fi\fi\fi\@esphack} \gdef\@eqnlabel{#1}}
   \def\@eqnlabel{} \def\@vacuum{}
   \def\draftmarginnote#1{\marginpar{\raggedright\scriptsize\tt#1}}

\def\draft{\oddsidemargin -.5truein \def\@oddfoot{\sl preliminary draft
	\hfil \rm\thepage\hfil\sl\today\quad\militarytime}
	\let\@evenfoot\@oddfoot \overfullrule 3pt \let\label=\draftlabel
	\let\marginnote=\draftmarginnote
	\def\@eqnnum{(\theequation)\rlap{\kern\marginparsep\tt\@eqnlabel}%
	\global\let\@eqnlabel\@vacuum} }

\def\numberbysection{\@addtoreset{equation}{section}
	\def\theequation{\thesection.\arabic{equation}}}

\def\underline#1{\relax\ifmmode\@@underline#1\else
	$\@@underline{\hbox{#1}}$\relax\fi}

\def\titlepage{\@restonecolfalse\if@twocolumn\@restonecoltrue\onecolumn
     \else \newpage \fi \thispagestyle{empty}\c@page\z@
     \def\thefootnote{\fnsymbol{footnote}} }

\def\endtitlepage{\if@restonecol\twocolumn \else \fi
	\def\thefootnote{\arabic{footnote}} \setcounter{footnote}{0}}
\relax \def\ie{\hbox{\it i.e.}}  
      \def\tr{\mathop{\rm tr}} \def\Tr{\mathop{\rm Tr}}
\def\beq{\begin{equation}} \def\eeq{\end{equation}}
\def\bea{\begin{eqnarray}} \def\eea{\end{eqnarray}} \def\bar{\overline}
\def\z{\bar {z}} \def\nn{\nonumber} \def\pa{\partial} \def\d{{\cal D}}
\def\LL{{\cal L}}

\def\P{\Psi} \def\Pb{{\bar{\Psi}}} \def\F{ \Phi} \def\Fb{{\bar{\Phi}}}
\def\etab{\bar{\eta}} \def\m{\mu} \def\n{\nu} \def\r{\rho} \def\s{\sigma}
\def\t{\tau} \def\m{\mu} \def\e{\epsilon} \def\k{\kappa} \def\kb{{\bar \k}}
\def\a{\alpha} \def\b{\beta} \def\g{\gamma} \def\d{\delta} \def\l{\lambda}
\def\demi{{1\over 2}} \def\f{\varphi} \def\fb{{\bar\varphi}} \def\v{v}
\def\vb{{\bar v}} \def\sb{{\bar s}} \def\sb{{\bar s}} \def\n{\eta}
\def\nb{{\bar \eta}} \def\xb{{\bar \kappa}} \def\Xi{{ \bar \kappa}}

\def\B{ BRST{$\ $} } \def\AB{ anti-BRST{$\ $} } \def\T{ Topological Quantum
Field Theory{$\ $} } \def\TS{Topological Quantum Field Theories{$\ $}}
\def\ba{{\bar\alpha}}

\def\ee{\eea} \def\be{\bea}

\def\eq{\eeq} \def\bq{\beq}

\def\fs{f^\a_{\b \g}} \def\fss{f^A_{B \g}} \def\cb{{\bar c}}\def\sb{\bar s}

\relax 

\numberbysection \hybrid
\begin{document}
\begin{titlepage}
\begin{center} \hfill PAR--LPTHE 95--04 \\[1in] {\large\bf On The
 Symmetries Of Topological Quantum Field Theories}\\[1in] {\bf Laurent
 Baulieu }\footnote{email address: baulieu@lpthe.jussieu.fr} \\ {\it
 LPTHE\/}\footnote{Laboratoire associ\'e No. 280 au CNRS}\\ \it
 Universit\'e Pierre et Marie Curie - PARIS VI\\ \it Universit\'e Denis
 Diderot - Paris VII\\ Boite 126, Tour 16, 1$^{\it er}$ \'etage \\ 4 place
 Jussieu\\ F-75252 Paris CEDEX 05, FRANCE\\

\end{center}

\vskip 1in

\begin{quotation} {\bf Abstract }$\ $ We display properties of the general
formalism which associates to any given gauge symmetry a topological action
and a system of topological BRST and anti-BRST equations.  We emphasize the
distinction between the antighosts of the geometrical BRST equations and
the antighosts occuring in field theory.  We propose a transmutation
mechanism between these objects. We illustrate our general presentation by
examples.\end{quotation} \end{titlepage} \newpage

\section{Introduction} It is now well known that any given classical gauge
 symmetry can be put in correspondence with a BRST and anti-BRST
 differential algebra \cite{ag}. This permits an elegant and systematic
 construction of quantum field theories with a Hilbert space describing the
 same number of gauge invariant degrees of freedom as the associated
 classical gauge theories.  In this approach, the understanding of the
 statistics of various ghost and Lagrange multiplier fields is
 straightforward.  The selection of physical observables, as well as the
 identification and the control of anomalies, amount to the algebraic
 problem of solving the cohomology of the BRST operator.  This general and
 aesthetical procedure is very powerful.  As an example, it permits one to
 handle in a strikingly unified way systems with degenerate gauge
 symmetries, that is systems which contain ghost of ghost fields at the
 quantum level with possibly higher order ghost interactions, and also
 systems with global zero modes.  It contains as a limiting case the
 Faddeev-Popov method.

In this paper, I will present properties concerning the generalization of
this "principle of BRST symmetry" for topological quantum field theories
\cite{review}.  The motivation is two -fold. Firstly, starting from a gauge
symmetry, one ends up with a beautiful differential algebra, which includes
as a "hard kernel " the usual BRST and anti-BRST transformations laws of
the gauge symmetry.  Secondly, this provides a general framework to
quantize lagrangians which can be written locally as pure divergences.
Indeed, inspired by Witten ideas \cite{wittendona}, physicists have
realized that it makes sense to quantize such lagrangians although they do
not generate classical equations of motion.  The idea is to introduce an
equal number of commuting and anticommuting fields, in order to get a
configuration space with an effective number of degrees of freedom almost
equal to zero, and to define the propagation of modes by BRST invariant
gauge fixing terms with appropriate compensations between fermionic and
bosonic degrees of freedom.  Eventually, it becomes possible to compute non
trivial topological information by path integrations of certain operators,
which count various well weighted combinations of contributions of zero
mode of Dirac-like operators.  Many examples show that the general
formalism presented in this paper is appropriate for such quantization
schemes \cite{review}.

We will also introduce a mechanism which involves more fields than the
usual BRST formalism, with the following motivation. In quantum field
theory, the ghost fields have a clear geometrical interpretation,
essentially because their quantum numbers are the same as those of the
infinitesimal gauge transformations.  The antighosts do not have such a
natural interpretation. Rather, they play the role of Lagrange multipliers
for the BRST variations of gauge functions and their quantum numbers are
thus generally different from those of the ghosts, in a gauge dependent
way.  On the other hand, in the geometrical construction of the BRST and
anti-BRST symmetry, one finds that geometrical antighosts can be introduced
on an equal footing as ghosts. This distinction between the geometrical
antighosts and the field theory antighosts can be seen as an unpleasant
feature, or even as a contradiction.  Our point of view is that it is the
signal that the BRST formalism should beimproved to permit a transmutation
mechanism between these fields in a framework were a full ghost-antighost
and BRST anti-BRST symmetry is pre-existing before the construction of
gauge-fixed BRST invariant lagrangians. We thus propose in this paper to
enlarge the set of fields of the BRST formalism: in addition to the
geometrical ghosts and antighosts which are the building blocks of the
gauge symmetry BRST and anti-BRST differential algebra, we introduce the
field theory antighosts as elements of pairs with trivial BRST and
anti-BRST cohomology. Then, the derivation of field theory actions which
have possible asymmetries in their ghost and antighost dependances is done
by suitable gauge-fixing terms which eliminate effectively some of the
fields by supersymmetric compensations.  This mechanism, where one starts
from a configuration space with a full symmetry between the ghosts and the
antighosts, can occur in ordinary gauge theories as well as in topological
field theories.

One could even speculate on further developments based on the possibility
of introducing ex-nihilo new fields which count locally as zero degrees of
freedom in a way which is compatible with a classical gauge symmetry.
Their BRST invariant coupling to ordinary gauge invariant non-topological
models could be a hint to define new type of order parameters and to
determine to which phase these systems belong, for instance a confined or
deconfined phase in Yang-Mills theories, through spontaneous breaking of
either the ghost number conservation symmetry or the topological BRST
symmetry. It could have applications also in the interpretation of the
Gribov problem in the BRST formalism \cite{fu}, \cite{zwan}.

The paper is organized as follows: we recall in a first section the BRST
 and anti-BRST formalism for general non-topological field gauge symmetries
 and explain in this case the transmutation mechanism. We give as an
 example the bosonic string theory, expressed in the conformal gauge (more
 precisely in gauges where the conformally invariant part of the metric is
 set equal to a background value). Then, we consider the case of
 topological field theories with an inner gauge symmetry for which we show
 that the BRST anti-BRST formalism and the transmutation mechanism can be
 also applied, in the antighost as well as in the antighost for antighost
 sectors. We emphasize the possibility of using the anti-BRST symmetry as a
 useful tool to select interesting gauge choices for these theories and
 give the Donaldson-Witten theory as an example.

 \section{ BRST and anti-BRST formalism associated to a gauge theory, and
 antighost transmutation mechanism} Let us consider a system of fields
 $\f^i(x)$, $1\leq i\leq N$ which undergo the following infinitesimal gauge
 transformations
\be
\label{gauge} \d_\e \f^i(x) =R^i_\a(\f^i(x)) \e^\a (x) \ee The local
parameters $\e^\a(x)$, $1\leq \a\leq r < N$, can be commuting and/or
anticommuting, and the $R^i_\a$ are functions of the fields $\f$. $x$
denotes the space-time variable. We assume the consistency of these gauge
transformations, that is their closure relation and (graded) Jaccobi
identity \be\label{closure} [\d_{\e} ,\d_{\e'}]\f^i
=f^{\alpha}_{\beta\gamma} R^i_\alpha\e^\beta{\e'}^\gamma \ee
\be\label{Jaccobi} \left ([[\d_{\e} ,\d_{\e'}],\d_{\e''}]+{ cyclic\ and/or\
anticyclic permutations}\right )\f^i =0 \ee The $\fs$'s can be functions of
the fields $\f$'s, since this situation can occur in physics.

As shown in \cite{ag}, one can construct two graded differential operators
 $s$ and $\bar s$, called the BRST and anti-BRST operators associated to
 the gauge symmetry defined in \ref{gauge}. $s$ and $\sb$ act on an
 enlarged set of fields $\f^i$, $c^\a$, $\bar c^\a$ and $B^\a$ with the
 property \be s^2=s\sb+\sb s=\sb^2=0\nn\\ sd+d s=\sb d+\sb d=0 \ee
 $d=dx^\m\pa_\m$ is the exterior derivative.  The ghosts and antighosts
 $c^\a$ and $\bar c^\a$ have the same quantum numbers as the local
 parameters $\e^\a(x)$ but the opposite statistics, while the fields $B^\a$
 have the same quantum numbers and statistics as the $\e^\a(x)$. One
 assigns ghost numbers $0$, $1$, $-1$ and $0$ to $\f^i$, $c^\a$, $\bar
 c^\a$ and $B^\a$ respectively. One defines the total grading of any given
 product of fields as the sum of their form degrees and ghost numbers. In
 the next section, devoted to topological field theories, we will refine
 this grading by a splitting along ghost and antighost directions.

It is convenient to set the basic fields on the following diagram
\be\label{gaugespectrum} &\f^i & \nn\\ c^\a&\quad& \cb^\a \nn\\ &B^\a & \ee
$s$ and $\bar s$ are defined as follows \be\label{ghost} s\f^i=R^i_\a c^\a
\quad & \bar s\f^i=R^i_\a \cb^\a \nn\\ sc^\a=-\demi\fs c^\b c^\g \quad &
\bar s\cb^\a=-\demi\fs \cb^\b \cb^\g \ee \be s\bar c^\a+\bar s b^\a+ \fs
\cb^\b c^\g=0 \ee One assumes that $s$ and $\bar s$ commute with the space
derivative $\pa_\m$, that is \be sd+d s=\sb d+\sb d=0 \ee Equations
\ref{closure} and \ref{Jaccobi} imply relations between the $\fs$ and the
$R^i_\a$ which are exactly what is needed to prove the property that
\be\label{nil} s^2=s\sb+\sb s=\sb^2=0 \ee on all fields $\f^i$, $c^\a$ and
$\bar c^\a$ , even in cases where the $\fs$ are field dependent \cite{ag}.

Notice that there is yet no $B$ dependence in the equations, and that
$s\cb^\a $ and $ \bar s B^\a$ are undetermined.  The introduction of $B$
through the following definition, raises this degeneracy, while maintaining
automatically \ref{nil} \be\label{bghost} s\cb^\a=B^\a\quad & \sb
c^\a=-B^\a- \fs B^\b \cb^\g \nn\\ s B^\a=0\quad & \bar s\bar B^\a= - \sb
\left(\fs B^\b \cb^\g\right) \ee One expects that the cohomology of the
operations $s$ (resp. $\sb$) with zero or positive (resp. negative) ghost
number only involves functions of the fields $\f$ and $c$
(resp. $\cb$). This remark would be important if one one were to
investigate the classification of possible anomalies of the gauge symmetry.

After having introduced the graded differential operators $s$ and $\bar s$
through this construction, (which can be quite easily generalized for
symmetries requiring ghosts of ghosts), one can solve the problem of gauge
fixing a lagrangian $\LL_{cl}(\f)$ invariant under the symmetry
\ref{gauge}. The principle is to add to $\LL_{cl}(\f)$ a BRST invariant
term which is BRST-exact and induces a propagation of the "longitudinal"
modes, that is the modes which can be gauged away and have therefore no
classical dynamics induced by $\LL_{cl}(\f)$. The quantum lagrangian is
therefore \be \LL_{cl}(\f)\to \LL_{Q}(\f,c,\cb,B)=\LL_{cl}(\f)+s(\bar
K_{-1}) +s\sb(K_0) \ee $ K_{-1}$ and $K_0$ should be well chosen local
functions of all fields: after expansion of $s(\bar K_{-1}) +s\sb(K_0)$ one
must get terms which are not gauge invariant and which determine a local
propagation of the longitudinal degrees of freedom of the fields. There is
less arbitrariness in their choices if one requires that the action has
ghost number zero, in which case $ K_{-1}$ and $K_0$ have ghost number -1
and 0 respectively, and also that it is invariant under relevant global
symmetries, such as the Lorentz invariance in particle models. Power
counting argument can be used too, and some gauge choices can possibly
present interesting accidental symmetries.

The BRST invariance of the action implies Ward identities which can be used
 to determine the whole quantum theory. Moreover they allow one to classify
 the possible anomalies as the solutions of consistency equations, and to
 determine their influence on the theory. The determination of the
 cohomology of the BRST Hamiltonian charge is a clear way of selecting the
 physical observables: they must commute with this charge and not be BRST
 exact.  It also permits one to separate the Hilbert space of the theory
 into physical and unphysical sectors. In practice, these very general
 features must of course be verified according to the computation rules
 specific to the model that one considers.  In particular, one must
 carefully determine the BRST charge and ghost number charge of the vacuum,
 in order to justify the above definition of the observables.

  The way of thinking which leads one to the BRST formalism and can be
  called "principle of BRST
symmetry" might appear as too abstract.  However, all along the road, the
physicist keeps a rather natural guide-line: the introduction of ghosts as
propagating fields is a necessity to compensate the propagation of
unphysical, \ie gauge dependent, components of the classical field in a
covariant way. Then, the BRST symmetry can be thought of as a natural
definition of the gauge symmetry for the enlarged set of fields including
the ghosts. Once these ideas becomes intuitive ones, one may raise at the
level of a principle the requirement of BRST symmetry.

  In this geometrical set-up, the following observation is however
  troublesome. Due to the freedom of
choosing different gauges, the quantum numbers of the antighosts and
Lagrange multipliers that one uses in quantum field theory do not generally
coincide with the quantum numbers of the gauge symmetry parameters. More
precisely, if one denotes by $\l^A $ the lagrange multipliers of the gauge
functions, then the indices $A$ and $a$ run over the same number of
independent values, but they generally describe different representation
spaces.  Therefore the Lagrange multipliers $\l^A $ should not be
identified with the fields $B^\a$ in \ref{bghost}, although the latter
enter so naturally in the geometrical construction of the BRST
symmetry. This situation occurs for instance in the quantization of the
bosonic string in the conformal gauge, for which $\a$ denotes a
2-dimensional vector field while $A$ denotes a 2-dimensional quadratic
differential.  In contrast, the example of the Yang-Mills in Lorentz type
gauges is simplest: $\l$ has in this case the same quantum numbers as the
infinitesimal parameters of a Yang-Mills transformation, and can thus be
identified with the geometrical field $B$.

 To reconcile the beauty of the geometrical construction of the BRST and
 anti-BRST symmetry with the
necessary freedom of the choice of the gauge functions, it is therefore
tempting enough to try to find a way of transmuting the pair $\cb^\a, B^\a$
into another pair $\Xi^A, \l^A$. As we will see shortly, a natural way of
doing this is the introduction of $\l^A $ as part of a quartet of fields
which count as a whole for zero degrees of freedom and undergo the BRST and
anti-BRST symmetry in a way which is cohomologically trivial.

  We thus add to the field spectrum \ref{gaugespectrum}
the following field quartet \be\label{new} &L^A & \nn\\ \eta^A&\quad& \xb^A
\nn\\ &\l^A & \ee One assigns ghost numbers $0$, $1$, $-1$ and $0$ to
$L^A$, $\n^A$, $\xb^A$ and $\l^A$ respectively, and one defines \be
sL^A=\n^A & \quad & \sb L^A=\xb^A \nn\\ s \n^A=0 &\quad & \sb \xb^A= 0\nn
\ee \be
\label{bghost1} s\xb^A+\sb \n^A=0 \ee These equations can be written under
the following form, which is useful in view of a comparison with the
transformation laws of a topological field theory \be
(s+\sb)L^A=\n^A+\xb^A\nn\\ (s+\sb)(\n^A+\xb^A)=0 \ee To determine $s\xb^A $
and $ \bar s \n^A$, while maintaining \ref{nil}, we set \be\label{bghost2}
s\xb^A=\l^A & \quad & \sb \n^A=-\l^A \nn\\ s \l^A=0 &\quad & \sb \l^A= 0
\ee If the index $A$ has a geometrical meaning, structure functions $\fss$
analogous to the $\fs$ should exist, with a Jaccobi identity of the type
\ref{Jaccobi}. Then, it is geometrically meaningful to redefine $\n^A \to
\n^A-\fss c^\g L^B$, $\Xi^A \to \Xi^A-\fss \cb^\g L^B$ and $\l^A \to
\l^A-\fss c^\g \Xi^B$. This amounts to the system \be\label{geoo}
(s+\sb)L^A+\fss(c^\g+\cb ^\g)L^B=\n^A+\Xi^A\nn\\
(s+\sb)(\n^A+\Xi^A)+\fss(c^\g+\cb ^\g)(\n^A+\Xi^A)=0 \ee We will check the
usefulness of these redefinitions on the bosonic string example.

To substitute the pairs $\cb,B$ into the pairs $\Xi,\l$ we will use the
form of the BRST symmetry for the fields in \ref{new} which permits, as we
will see, the decoupling of certain fields from the lagrangian without
changing the effective number of degrees of freedom of the theory.  To
explain the mechanism it is sufficient to consider the definition of the
BRST symmetry in \ref{bghost1} rather than in \ref{geoo}. We assume the
existence of possibly field dependent operators $O_{A\a}$ which relate the
representation spaces denoted respectively by the indices $A$ and $\a$.
Then, the possibility of a BRST invariant transmutation relies on the
following identity \be\label{identity} \int
[d\f][dc][d\cb][dB][dL][d\n][d\Xi][d\l]\exp S[\f,c,\Xi,\l] +\int dx
s(\cb^\a O_{A\a}L^A)\nn\\ \sim \int [d\f][dc][d\Xi][d\l]\exp S[\f,c,\Xi,\l]
\ee Indeed, the result of the path integration of $\exp \int dx s(\cb^\a
O_{A\a}L^A)$ over the fields $B,L,\cb,\n$ is formally equal to one, as a a
ratio of equal determinants, since one has \be s(\cb^\a O_{A\a}L^A)=
(B^\a-\cb^\b sO_{B\b} O^{-1 \a B})O_{A\a}L^A-\cb^\a O_{A\a}\n^A \ee

 The effect of the BRST invariant gauge fixing term $s(\cb^\a
O_{A\a}L^A)$ is thus an effective decoupling of the modes contained in the
fields $\cb,B,L,\n$. The remaining action $S[\f,c,\Xi,\l]$ is generally
dissymmetric in the ghosts and antighosts since it involves only
sub-sectors of the quartets of fields defined in \ref{gaugespectrum} and
\ref{new}.

 Other possibilities exist: instead of $s(\cb^\a O_{A\a}L^A)$, one could
consider expressions of the type $s\sb (\f^i O_{i\a}L^A)$ which give the
anti-BRST invariance in an automatic way. Once again, let us stress that in
the context of field theory, the validity of \ref{identity} should be
verified case by case, according to the computation rules of the models
that one wishes to explore.

The advantage of maintaining the anti-BRST symmetry is that this symmetry
can be used as a tool to examine the possible background gauge invariances
of the model. Let us suppose indeed that $S[\f,c,\Xi,\l]$ is $s$ and $\sb$
invariant.  After the transmutation, the remaining fields undergo the
following transformations \be\label{rghost} s\f^i=R^i_\a c^\a \quad & \bar
s\f^i=R^i_\a \cb^\a \nn\\ sc^\a=-\demi\fs c^\b c^\g \quad & \sb c^\a= -\fs
c^\b \cb^\g \nn\\ s\xb^A=\l^A-\fss c^\g \Xi^B \quad & \sb \xb^A= -\fss
\cb^\g \xb^B \nn\\ s \l^A=-\fss c^\g \l^B \quad & \sb \l^A= -\fss \cb^\g
\l^B\ee The action is independent on $\cb^\a$ which, from the definition of
$\sb$, can be understood as the ghostified parameter of a background
symmetry.

\def\z{{\bar z}} \def\m{\mu ^z_{\z}} \def\mb{{\mu _z^{\z}}} \def\c{c ^z }
\def\cb{{\bar c} ^z } \def\B{B ^z } \def\L{L_{zz} } \def\n{\eta_{zz} }
\def\Xi{{\bar \kappa}_{zz} } \def\xb{\kappa_{zz} } \def\l{\lambda_{zz}}
\def\b{b_{zz} } \def\d{{\pa} _z } \def\db{{\pa} _\z } \def\D{{\db -\m\d}}
\def\DB{{\d -\mb\db}} \def\mm{1-\mu ^z_{\z} \mu ^{\z}_z} \def\Dc{ \db\c
+\c\d\m-\m\d\c} \def\Dcb{ \db\cb +\cb\d\m-\m\d\cb} \def\DXi{ \db\c
+\c\d\m-\m\d\c} \def\DXi{ \d\cb +\cb\d\m-\m\d\cb} \def\mo{\mu ^z_{0\z}} To
illustrate these general formula, let us consider the bosonic string
theory. Using the Beltrami differential parametrization, the classical
string lagrangian is \cite{bellon} \be {\it L}_{string}={{1}\over{\mm}}
(\D) X(\DB )X \ee The role of $\f$ is played by the Beltrami differential
$\m$.  The BRST equations in the $2-D$ gravity sector are
\be\label{rghostbel} s\m=\Dc \quad & &\bar s\m=\Dcb \nn\\ s\c=\c\d\c \quad
& &\sb\cb=\cb\d\cb \nn\\ s\cb=\B \quad & &\sb \c=-\B-\c\d\cb-\cb\d\c \nn\\
s \B=0 \quad & &\sb \B= -\cb\d\B+\B\d\cb \ee \be\label{rghostbelll}
s\L=\n-\c\d\L+2\L\c \quad & &\bar s\L=\Xi-\cb\d\L+2\L\d\cb\nn\\
s\n=\c\d\n+2\n\d\c \quad & &\sb \Xi=-\cb\d\Xi+2\Xi\d\cb \nn\\ s\Xi=\l
-\c\d\Xi-2\Xi\d\c \quad & &\sb \n=-\l-\cb\d\n-2\n\d\cb \nn\\ s
\l=-\c\d\l+2\l\d\c \quad & &\sb \l= -\cb\d\l+\l\d\cb \ee One has of course
the mirror equation in the antiholomorphic sector, by changing $z\to\z$ and
$\z\to z$.  To define the theory in the gauge where the Beltrami
differential $\m$ is equal to a background value $\mo$, we use our general
formalism and add to the string action the following BRST invariant term
\be {\it L}_{gf}=s\sb\left(\L(\m-\mo)\right)+c.c \ee This yields the usual
result of the conventional BRST quantization. Indeed, an easy computation
gives firstly \def\bet{\beta_{zz}} \be L_{gf}=s\left(\Xi'(\m-\mo)-\L(\Dcb
)\right) \ee and then \be L_{gf}={\l}' (\m-\mo) -\b(\Dc) \nn\\
+{\n}'(\D+\pa_z\m)\cb \nn\\ -{\L}'(\D+\pa_z\m)\B+c.c \ee We have done
simple field redefinitions ${\l}' =\l+\ldots$, ${\n}'=\n+\ldots$,
${\L}'=\L+\ldots$, $\b=\Xi+\ldots$.  \def\l{\bet} The two last of $ L_{gf}$
compensate each other in the path integration by N=2 supersymmetry.
Curiously, they can be understood as a particular case of a bidimensional
topological gravity.

The two remaining terms are the known gauge fixing terms of the bosonic
string theory, so we have in illustration of the identity \ref{identity},
with \be S\sim\int dzd\z ({{1}\over{\mm}} (\D) X(\DB )X \nn\\-\b(\Dc)-c.c )
\ee Moreover, the BRST anti-BRST symmetry operators $s$ and $\sb$ which
leave invariant this action and satisfy $s^2=s\sb +\sb s=\sb^2=0$ are
\be\label{rghostbbbbbb} s\m=\Dc \quad & &\bar s\m=\Dcb \nn\\ s\c=\c\d\c
\quad & &{\sb}\cb=\cb\db\cb \nn\\ s\cb=0 \quad & &\sb \c=-\c\d\cb-\cb\d\c
\nn\\ s\b=0 \quad & &\sb \b=-\cb\d\b-2\b\d\cb \ee $\sb$ can be interpreted
as the ghostified form of the background diffeormorphism symmetry, with the
spectator $2-D$ vector field antighost, and can be used as in \cite{bellon}
to deduce the "Virasoro Ward identity" of the conformal theory.

\def\z{\bar {z}} \def\nn{\nonumber} \def\pa{\partial} \def\d{{\cal D}}
\def\LL{{\cal L}}

\def\P{\Psi} \def\Pb{{\bar{\Psi}}} \def\F{ \Phi} \def\Fb{{\bar{\Phi}}}
\def\etab{\bar{\eta}} \def\m{\mu} \def\n{\nu} \def\r{\rho} \def\s{\sigma}
\def\t{\tau} \def\m{\mu} \def\e{\epsilon} \def\k{\kappa} \def\kb{{\bar \k}}
\def\a{\alpha} \def\b{\beta} \def\g{\gamma} \def\d{\delta} \def\l{\lambda}
\def\demi{{1\over 2}} \def\f{\varphi} \def\fb{{\bar\varphi}} \def\v{v}
\def\vb{{\bar v}} \def\sb{{\bar s}} \def\sb{{\bar s}} \def\n{\eta}
\def\nb{{\bar \eta}} \def\xb{{\bar \kappa}} \def\Xi{{\bar \kappa}}

For cases where the gauge function has the same quantum numbers as the
 parameters of the theory, (for instance the Yang-Mills theory with a
 Lorentz gauge function), there is no need to introduce the quartet of
 fields \ref{new}. Otherwise, these fields can be fully decoupled by mean
 of the following action \be s(\xb^A L^A)=\l^A L^A -\xb^A \n^A \ee

\def\tr{\mathop{\rm tr}} \def\Tr{\mathop{\rm Tr}}
\def\beq{\begin{equation}} \def\eeq{\end{equation}}
\def\bea{\begin{eqnarray}} \def\eea{\end{eqnarray}} \def\bar{\overline}
\def\z{\bar {z}} \def\nn{\nonumber} \def\pa{\partial} \def\d{{\cal D}}
\def\LL{{\cal L}}

\def\P{\Psi} \def\Pb{{\bar{\Psi}}} \def\F{ \Phi} \def\Fb{{\bar{\Phi}}}
\def\etab{\bar{\eta}} \def\a{\alpha} \def\m{\mu} \def\n{\nu} \def\r{\rho}
\def\s{\sigma} \def\t{\tau} \def\m{\mu} \def\e{\epsilon} \def\k{\kappa}
\def\kb{{\bar \k}} \def\a{\alpha} \def\b{\beta} \def\g{\gamma}
\def\d{\delta} \def\l{\lambda} \def\demi{{1\over 2}} \def\f{\varphi}
\def\fb{{\bar\varphi}} \def\v{v} \def\vb{{\bar v}} \def\sb{{\bar s}}
\def\sb{{\bar s}} \def\n{\eta} \def\nb{{\bar \eta}} \def\xb{{\bar \kappa}}
\def\Xi{{\bar \kappa}} \def\o{\omega}

\def\B{ BRST{$\ $} } \def\AB{ anti-BRST{$\ $} } \def\T{ Topological Quantum
Field Theory{$\ $} } \def\TS{Topological Quantum Field Theories{$\ $}}
\def\ba{{\bar\alpha}}

\def\ee{\eea} \def\be{\bea}

\def\eq{\eeq} \def\bq{\beq}

\def\fs{f^\a_{\b \g}} \def\fss{f^A_{B \g}} \def\cb{{\bar c}}\def\sb{\bar s}
\section{Topological BRST and anti-BRST equations associated to a gauge
theory}

We will now consider the case of topological field theories.  We consider
the same classical fields as in the previous section, but assume that they
are submitted to arbitrary local fields transformations.  Since the
geometrical idea is to describe such generalized gauge transformations
modulo the usual gauge transformations \ref{gauge}, we introduce the
following infinitesimal transformations \be \label{gaugetop} \d_\e \f^i
=\e^i(x) +R^i_\a \e^\a (x) \ee The local parameters $\e^\a(x)$ have the
same meaning as in \ref{gauge} while the $\e^i(x)$ have the same quantum
numbers as the fields $\f^i$'s.

One has obviously a "gauge invariance" in the space of transformation
parameters \be \e^\a (x)\to \e^\a (x)+\o^\a (x) \nn\\ \e^i (x)\to \e^i
(x)-R^i_\a \o^\a (x) \ee

The symmetry \ref{gaugetop} is an invariance of topological terms, if any,
which can be expressed as functionals of the $\f$'s. It represents the
largest gauge symmetry acting on the $\f$'s. To build its associated graded
differential BRST and anti-BRST operations, we must introduce more fields
than in \ref{gaugespectrum}. We consider the following set of fields
\be\label{spectrum} & \f^{i (0,0)} & \nn\\ \P^{i (1,0)} \quad & B^{i (1,1)}
& \quad{\Pb }^{i (0,1)} \nn\\ \F^{\a (2,0)} \quad\quad \eta^{\a(1,0)} \quad
& L^{\a (1,1)} &\quad\bar{\eta}^{\a (0,1)} \quad\quad \Fb^{\a (0,2)}\nn\\
c^{\a (1,0)} \quad & B ^{\a (1,1)} & \quad{\cb} ^{\a (0,1)} \ee Notice that
we have improved the ghost number attributions of fields: the integer
indices $g$ and $\bar g$ count respectively the ghost and antighost numbers
of a field $X^{g,\bar g}(x)$ and determine a bi-grading. The knowledge of
the sum of the form degree and of $g-\bar g$ is however sufficient to
determine the commutation properties: if this quantity is even (resp. odd),
$X^{g,\bar g}$ has the physical (resp. unphysical) spin-statistics
relation.

The topological BRST and anti-BRST operators $s$ and $\bar s$ associated to
\ref{gaugetop} are defined as follows (we use the notation that $X_{,i}=
{{\d X}\over {\d\f^i}}$): \be\label{ghosttop} s\f^i = \P^i+R^i_\a c^\a\quad
& \bar s\f^i = \Pb^i+R^i_\a \cb^a \nn\\ sc^\a = \F^\a-\demi\fs c^\b
c^\g\quad & \bar s\cb^\a = \Fb^\a-\demi\fs \cb^\b \cb^\g \nn\\ s\P^i =
-R^i_\a \F^\a-R^i_{,j \ \a} c^\a\P^j \quad & \sb \Pb^i = -R^i_\a
\Fb^\a-R^i_{,j \ \a}\cb^\a \Pb^j \nn\\ s\F^\a = \fs \F^\g c ^\b +\demi
f^\a_{,i\b\g }\P^i c^\b c^\g\quad & \bar s\Fb^\a = \fs \Fb^\g c ^\b +\demi
f^\a_{,i\b\g } \Pb^i \cb^\b \cb^\g\nn\\ \ee and

\be \label{mixedtop} s\bar c^\a+\bar sc^\a+ \fs \cb^\b c^\g+L^\a=0 \nn\\
s\Pb^i+\sb \P^i +R^i_\a L^\a+R^i_{,j \ \a}\Pb^j c^\a+ R^i_{,j \ \a}\P^j
\cb^\a=0 \nn\\ s \Fb^\a+\sb L^\a+\fs c ^\b \Fb^\g+\fs \cb ^\b L^\g +\demi
f^\a_{,i\b\g } \P^i \cb^\b \cb^\g=0 \nn\\ \sb \F^\a+s L^\a+\fs \cb ^\b
F^\g+\fs c ^\b L^\g +\demi f^\a_{,i\b\g } \Pb^i c^\b c^\g=0 \ee As for the
pure gauge transformation BRST algebra, the closure and Jaccobi relations
\ref{closure} and \ref{Jaccobi} of the gauge symmetry enforce the
fundamental nilpotency property $s^2=s\sb+\sb s=\sb^2=0$ on all fields.
The topological BRST equations yield a consistent separation between
arbitrary fields deformations and the pure gauge transformations
\ref{gauge}: the latter can be seen as a meaningful truncation of the
latter. Observe that the idea of having this kind of field spectrum for a
topological symmetry with a BRST and anti-BRST symmetry as in
\ref{ghosttop} was first introduced in the particular context of the
Yang-Mills theory \cite{stto}.

The geometrical equations \ref{mixedtop} mix ghost and antighost parts.
Indeed, these equations could be obtained directly from the sole definition
of the operation $s$ on $\f$, $c$, $\P$, $\F$ by changing $s\to s+\sb$,
$c\to c+\cb $, $\P\to \P+\Pb$, $\F\to \F+L+\Fb$. To solve the degeneracy in
\ref{mixedtop}, one uses the "auxiliary" fields $B$, $\P$, $\eta$ and $\bar
\eta$ in \ref{spectrum}.

Firstly, one defines \be\label{aux} s\cb^\a=B^\a \quad & sB^\a=0 \nn\\
s\Pb^i=B^i \quad & sB^i=0 \nn\\ s\Fb^\a=\etab^\a \quad & s \etab^\a=0 \nn\\
sL^\a= \eta ^\a \quad & s\eta^\a=0 \ee Then, the rest of the equations is
defined as follows: one obtains the action of $\sb$ on $c$, $\P$, $L$ and
$\F$ by combining the last equation and \ref{mixedtop}; by imposing that
$\sb ^2 =0$ on these fields, one finds afterwards the action of $\sb$ on
the auxiliary fields $B$, $\P$, $\eta$ and $\bar \eta$; finally, one
obtains the property $(d+s+\sb)^2=0$, which means that $d$ anticommutes
with $s$ and $\sb$, and that $s^2=s\sb+\sb s=\sb ^2=0$ on all fields.

It is impossible to construct a local action invariant under the
transformations \ref{gaugetop} which would be only function of the
classical fields $\f$ and would generate equations of motions.  Indeed, the
gauge symmetry is so large that the only possibility would be to consider
lagrangians which are locally pure derivatives, which means that the action
is a topological invariant.  As a matter of fact, to construct a
topological field theory associated to this huge symmetry, one must define
the path integral by using fields which count as a whole as zero degrees of
freedom, which is the number of physical local degrees of freedom left
locally by the symmetry \ref{gaugetop}. Therefore, it is natural to
introduce the set of fields \ref{spectrum} as fundamental fields, and to
define the theory by the gauge functions associated to a lagrangian of the
following form \be \LL_{Q} =d\omega+s(\bar K_{-1}) +s\sb(K_0) \ee In other
words, our principle is to postulate the BRST invariance associated to the
symmetry \ref{gaugetop}.

 At this point, it is clear that we must face the fact that, as in ordinary
 gauge theories,
interesting gauge functions can take their values in various representation
 spaces, so that some of the geometrical fields displayed in
 \ref{spectrum}.  must be transmuted into field theory objects with
 different quantum numbers.  We will therefore generalize the idea
 explained in the previous section and couple the theory to cohomologically
 trivial pairs of additional pairs.

To display the formalism, we will firstly consider the cases for which the
 gauge functions for the "longitudinal" part of the fields $\f$ have the
 same quantum numbers as the gauge symmetry parameters. In this situation,
 there is no need of transmutation of the geometrical pair $\cb^\a,
 B^\a$. On the other hand, to gauge-fix the remaining $N-r$ "transverse"
 part of the fields $\f$, one must get rid of the pairs $\Pb^i$, $B^i$,
 $L^\a$ and $\n^\a$, and replace them by a pair $\xb^{M}, \l^{M}$, where
 the index $M$ runs between $1$ and $N-r$, to get eventually a gauge fixing
 term of the form $\l^M {\it {G}}_M(\f)+\ldots$.

 For this purpose, we introduce cohomologically trivial pairs with the
 relevant quantum numbers (denoted by the indices $M$, $N$, ...)
\be & L^{M (0,0)} & \nn\\ \eta^{M (0,0)}&\quad& \xb^{M (0,0)} \nn\\ &\l^{M
  (0,0)} & \ee Notice that the effective number of degrees of freedom
  carried by the fields $\Pb^i$, $B^i$, $L^\a$, $\n^a$ and $ L^M, \n^M$
  counts overall for zero.

To obtain a BRST invariant elimination of the fields $\Pb^i, B^i$, we
 include in the action a term of the following form \be\label{elimination}
 s\left( \Pb^i(O_{iM }L^M+O_{i\a }L^\a \right)=- \Pb^i(O_{iM }\n^M+O_{i\a
 }\n^\a ) \nn\\ (B^i-\Pb^j (sO_{j N })O^{-1Ni})O_{iM}L^A +(B^i-\Pb^j (sO_{j
 \b }) O^{-1\b i})O_{i\a}L^\a \ee Indeed, if one chooses appropriately the
 transfer operators $O_{iA }$ and $O_{i\a }$, one expects that the fields
 $\Pb^i, B^i$ and $L^A, L^\a, \n^A, \n^\a$ will decouple from the theory
 through supersymmetric compensations analogous to those which justify the
 identity \ref{identity}.

Let us stress that the presence of the "medium" geometrical ghost of ghost
 $L^\a$ is necessary to ensure automatically the relation $(s+\sb)^2=0$.
 Its elimination from the spectrum is thus generally not compatible with
 the existence of an operation $\sb$ which would anticommute with
 $s$. However, when the structure functions $\fs$ are field independent,
 which covers the case of all the gauge symmetries with a Lie algebra
 structure, a consistent $\sb$ operation still exists. Indeed, let us write
 the form of the BRST anti-BRST algebra which remains after the elimination
 of the fields $\Pb$,$L$ and $\n$: \be\label{ghosttopym} s\f^i=\P^i+R^i_\a
 c^\a \quad & \bar s\f^i= R^i_\a \cb^\a \nn\\ sc^\a=\F^\a-\demi\fs c^\b
 c^\g \quad & \bar s\cb^\a= -\demi\fs \cb^\b \cb^\g \nn\\ s\P^i=-R^i_\a
 \F^\a-R^i_{,j \ \a} c^\a\P^j \quad & \sb \P^i=-R^i_{,j \ \a}\cb^\a \Pb^j
 \nn\\ s\F^\a= \fs \F^\g c ^\b +\demi f^\a_{,i\b\g }\P^i c^\b c^\g \quad &
 \sb \F^\a=-\fs \cb ^\b \F^\g \nn\\ s \Fb^\a=\nb^a-\fs c ^\b \Fb^\g \quad &
 \bar s\Fb^\a= \fs \Fb^\g c ^\b \nn\\ s\nb^\a=\fs\F^\b\Fb^\g-\fs
 c^\b\nb^\g\quad & \sb\nb^\a= -\fs \cb^\b\nb^\g \ee \be \label{mixedtopym}
 s\bar c^\a+\bar sc^\a+ \fs \cb^\b c=0 \ee \be s \Xi^M=\l^M-f^M_{N\a}c^\a
 \Xi^N \quad & \sb \Xi^M= -f^M_{N\a}c^\a \Xi^N \nn\\ s \l
 ^M=-f^M_{N\a}\F^\a \Xi^N -f^M_{N\a}c^\a \Xi^N-f^M_{,jN\a}\P^j\F^\a \l^N
 \quad & \sb \l^M= -f^M_{N\a}\cb^\a \l^N \ee One can verify that the
 property $s^2=\sb ^2=0$ is still satisfied. However, the property that $s$
 and $\sb$ anticommute, that is $s\sb+\sb s=0$, is generally broken by
 terms proportional to the field derivatives of structure functions $\fs$
 and $f^M_{N\a}$.

To go further and enforce the anticommutation relation between $s$ and
$\sb$, we restrict therefore ourself to cases where these structure
functions are independent of the fields. Then, one can adopt a convenient
bracket anti-bracket notation, and write the BRST and anti-BRST equations
as follows \be\label{ghosttopbb} s\f =\P +R c \quad & \bar s\f = R \cb
\nn\\ sc =\F -\demi[c,c]\quad & \bar s\cb = -\demi[\cb,\cb] \nn\\ s\P =-R
\F -[c,\P] \quad & \sb \P =[\cb,\P]\nn\\ s\F= -[c,\F] \quad & \sb \F
=-[\cb,\F] \nn\\ s \Fb=\nb -[c,\Fb] \quad & \sb\Fb = [\cb,\Fb] \nn\\ s
\nb=[\F,\Fb] -[c,\nb] \quad & \sb\nb = -[\cb,\nb] \ee \be \label{ymm} s\bar
c +\bar sc +[c,\cb]=0 \ee \be s \Xi =\l -[c,\Xi] \quad & \sb \Xi=-[\cb,\Xi]
\nn\\ s \l =[\F,\Xi]-[c,\l] \quad & \sb \l = -[\cb,\l] \ee Notice that, in
the ghost sector, one can summarize all equations as follows \be
(s+\sb)(c+\cb)+\demi[c+\cb,c+\cb]=\F \nn\\ (s+\sb) \Xi + [c+\cb,\Xi]=\l
\nn\\ (s+\sb) \Fb + [c+\cb,\Fb]=\nb \ee with the "Bianchi identities" \be
(s+\sb)\F+ [c+\cb,\F]=0 \nn\\ (s+\sb)\nb+ [c+\cb,\nb]=[\F,\nb] \nn\\
(s+\sb)\l+ [c+\cb,\l]=[\F,\l] \ee In this way of writing the BRST equation,
the geometrical interpretation of the ghost of ghost $\F $ as the component
of a curvature \cite{bs} is almost obvious, since we have generally \be [
S, S ]=\F \ee where $S=s+ [ c, \ ]$.

One may wonder what is the role left to the anti-BRST symmetry after the
 elimination of the geometrical fields which break the symmetry between $s$
 and $\sb$. As a matter of fact, the last equations show that $\sb$ can be
 interpreted as a spectator field which "ghostifies" a background symmetry,
 in a way which generalizes the case of non-topological field theories.
 This permits the attribution of well-defined quantum numbers to all
 fields, including the topological ghosts and antighosts.

Let us now show that the anti-BRST symmetry permits one to distinguish
between the gauge fixing of the transverse and longitudinal modes.  On the
one hand, one observes that since the $\sb$ transformations of the pairs
$(\Xi^M, \l^M)$ and $(\F^\a, \n^\a)$ are analogous to ordinary gauge
transformations, having a $\sb$-invariant gauge fixing lagrangian of the
following form \be\label{transversefg} s(\Xi^M {\it {G}}_M(\f)+\Fb^\a{\it
{F}}_{\a i}(\f)\P^i) \ee implies that the gauge functions $\it {G}_M(\f)$
and ${\it {F}}_{\a i}(\f)\P^i$ for the "transverse" part of the gauge field
and for the degenerate topological ghost $\P^i$ are gauge covariant.  One
then notices that, due to the elimination of the field $L$, the field
functionals $\Xi^M {\it {G}}_M(\f)+\F^\a{\it {\Fb}}_{Mi}(\f)\P^i$ is
$\sb$-invariant without being $\sb$-exact. In contrast, the gauge fixing
part of the "longitudinal' modes can always be done by terms of the type
$s\sb (\f^i O_{ij} f^j+\ldots)$.

This suggests therefore that the search of the gauge fixing functions in
the transverse sector amounts to find the cohomology with ghost number -1
of the $\sb$ operation, while the gauge fixing terms for the longitudinal
sector belong to the trivial part.

We can summarize these remarks by writing the gauge fixing terms under the
following form \be \label{truc} \LL_{GF}\sim s(\Xi^M {\it
{G}}_M(\f)+\Fb^\a{\it {F}}_{\a i}(\f)\P^i) +\demi s\sb (\f^i O_{ij}
f^j+\ldots) \ee

It is nevertheless interesting enough that actions of the type
 \ref{transversefg} can be deduced from the action \be s\sb(L^M{\it
 {G}}_M(\f)+\cb^\a{\it {F}}_{\a i}(\f)\P^i) \ee provided that one uses as
 constraints the algebraic equations of motion stemming from the lagrangian
 \ref{elimination}.  The elimination of geometrical fields as $L$ and $\Pb$
 has thus induced a non trivial cohomology for the $\sb$ operation, which
 we believe useful for the classification of "interesting" gauge choices.

To make more explicit these observations, it is time to give an example.
  Let us chose the $4-D$ topological Yang-Mills theory, expressed in a
  Lorentz type gauge with self-duality gauge conditions \cite{wittendona}
  \cite{bs}.  In this case, $\f$ stands for the Yang-mills field $A_\m$,
  valued in a Lie algebra $\it{ G}$ and $R$ is the covariant derivative $
  D_\m$, $[\ ,\ ]$. The index $M$ means that $\Xi$ and $\l$ are Lie algebra
  valued self-dual 2-forms.  In what follows, products of two fields mean
  their trace in $\it{ G}$, and $[X,Y]$ is the graded commutator of $X$ and
  $Y$.

The BRST and anti-BRST equations are \be\label{ymmmm} sA_\m =\P _\m+D_\m c
\quad & \bar sA_\m = D_\m \cb \nn\\ sc =\F -\demi[c,c]\quad & \bar s\cb =
-\demi[\cb,\cb] \nn\\ s\P_\m =-D_\m \F -[c,\P_\m] \quad & \sb \P_\m
=-[\cb,\P_\m]\nn\\ s\F= -[c,\F] \quad & \sb \F =-[\cb,\F] \nn\\ s \Fb=\nb
-[c,\Fb] \quad & \bar s\Fb = [\cb,\Fb] \nn\\ s \nb=[\F,\Fb] -[c,\nb] \quad
& \sb\nb = -[\cb,\nb] \ee \be \label{mixedtopymmmm} s\bar c +\bar sc
+[c,\cb]=0 \ee \be s \Xi_{\m\nu} =\l_{\m\nu} -[c,\Xi_{\m\nu}] \quad & \sb
\Xi_{\m\nu}=-[\cb,\Xi_{\m\nu}] \nn\\ s \l_{\m\nu}
=[\F,\Xi_{\m\nu}]-[c,\l_{\m\nu}] \quad & \sb \l_{\m\nu} = -[\cb,\l_{\m\nu}]
\ee \be s \cb =B \quad & \sb c=-B-[\cb,c] \nn\\ s B =0 \quad & \sb B =
-[\cb,B] \ee A lagrangian which satisfies the above mentioned criteria of
$s$ and $\sb$ invariances, and is renormalizable by power counting is
\def\n{\nu} \be {\it{L}}=s\left( \Pb_i L^{0i}+\Pb_0 L \right) \nn\\ s\sb
\left( L^{\m\n}(F_{\m\n} +^*F_{\m\n}) +\demi \Pb_\m\P^\m +\demi A^\m A_\m
\right) \ee After elimination of the fields $\Pb_\m, B_\m, L_{\m\nu}, L,
\eta_{\m\nu}$ and $\eta$, the lagrangian has the form \be{\it{L}}\sim
s\left(\Xi^{\m\n}(F_{\m\n} +^*F_{\m\n}+\demi\l_{\m\n})+\Fb
D_\m\P^\m)\right) + s\sb\left(\demi A^\m A_\m \right) \ee

On this example, we check that the gauge fixing term for the transverse
 part of the gauge field belongs to the non trivial part of the cohomology
 with ghost number -1 of $\sb$, since $\Xi^{\m\n}(F_{\m\n}
 +^*F_{\m\n}+\demi\l_{\m\n})+\Fb D_\m\P^\m$ is $\sb$-invariant but not
 $\sb$-exact, while the gauge fixing term for the longitudinal part belongs
 to the trivial part, since it is $\sb$-exact \be s\sb\left(\demi A^\m A_\m
 \right)= s(A^\m\pa_\m = -\sb(A^\m\pa_\m c) =A^\m\pa_\m B+D^\m v\pa_\m \cb
 \ee \def\L{{ \Lambda}} \def\G {{ \Gamma}} \def\Gb{{\bar \Gamma}}
 \def\Y{{Y}} \def\Xb{{\bar X}} \def\X{{ X}} \def\N{{ N}} \def\Nb{{\bar N}}

A next level of complication of the transmutation mechanism is for theories
of the type of the conformal $2D-$topological gravity: in such cases one
should transmute also the antighost $\cb^\a$.  This could be done by a mere
combination of the results of this section and of the previous section, and
it is not worth giving the details.  Rather, let us concentrate on a more
exotic situation, although we are not aware of models for which it would
occur, when the gauge fixing functions ${\it {\it F}}_{\P}(\f)\P^i$ of the
degenerate topological ghosts $\P$ would have different quantum numbers
than those of the secondary antighosts $\Fb^\a$.  We will show that our
idea of a transmutation could be applied in this case.

 According to our point of view, we would need to introduce new
fields to transmute the pairs $\Fb^\a,\nb^\a$ into new pairs $\Xb,\Nb $
with the relevant quantum numbers.  After a little bit of experimentation,
one finds that one must introduce the following set of fields, such that
the addition of their degrees of freedom is once more effectively zero
\be\label{spectrumf} \G^{P (1,0)} \quad &\L^{P (1,1)} & \quad{\Gb }^{P
(0,1)} \nn\\ \X^{P(2,0)} \quad\quad \N^{P(1,0)} \quad & \Y^{P (1,1)} &\quad
{\Nb}^{P (0,1)} \quad\quad \Xb^{P (0,2)} \ee We are now familiar with the
most direct way to get the BRST and anti-BRST equations for such
fields. They are \be (s+\sb)(\G+\Gb)=\X+\Y+\Xb \nn\\ (s+\sb)(\X+\Y+\Xb)=0
\ee After expansion, one gets \be s \G=\X&\quad\sb \G =-\L+\Y \nn\\
s\Gb=\L&\quad\sb \Gb =\Xb \nn\\ s\X =0 &\quad\sb \X=\N \nn\\ s\Xb =\Nb
&\quad\sb \X=0 \nn\\ s\Y=-\N &\quad\sb \Y=-\Nb \nn\\ s\L =0 &\quad\sb
\L=-\Nb \nn\\ s\N=0 &\quad\sb \N=0 \nn\\ s\Nb=0 &\quad\sb \Nb=0 \ee
(Whenever the index $P$ has a geometrical meaning, one has the option of
redefining $\X+\Y+\Xb+[c+\cb, \X+\Y+\Xb]$.)

The transmutation mechanism is now quite simple: one considers the
following $s$-exact lagrangians \be s(\Fb^\a O_{\a P}\G^P)=\Fb^\a O_{\a
P}\X^P +(\nb^\a +\Fb^\beta sO_{\beta Q}O^{-1 Q\alpha})O_{\a P}\G^P \nn\\
s(Y^Q O_{Q P}{\Gb}^P)=Y^Q O_{Q P}\Lambda^P +(N^Q +Y^R sO_{RS}O^{-1
SQ})O_{QP}\G^P \ee These terms permit supersymmetric compensations between
the fields $\Fb^\a$, $\nb^\a$, $\X^P$, $\G^P$, $Y^P $, $\G^P$, $N^P$ and
$\Lambda^P$.  If the transition functions $O_{\a P}$ and $O_{Q P}$ are well
chosen, one can assume that these fields eventually decouple. Then, with
the remaining fields $\Xb^P$ and $\Nb^P$, one can perform the gauge fixing
of the geometrical topological ghosts \be s( \Xb^P{\it {\it
F}}_{Pi}(\f)\P^i) =\Xb^P{\it {\it F}}_{Pi}(\f)(\P^i+R^i_\a c^\a)\nn\\ +(
{\Nb}^P +\Xb^Q s{ {\it F}}_{Qj} O^{-1jP}) O_{Pi}\P^i \ee where the gauge
functions ${\it {\it F}}_{Pi}(\f)\P^i$ are now chosen at will. This
transmutation mechanism in the ghost of ghost sector is of course very
analogous to the one in the primary ghost sector.

\section{Conclusion} We have established general formula which associate to
any given system of gauge transformation a system of topological BRST and
anti-BRST equations. By coupling these equations to additional pairs of
fields with trivial cohomology one can in principle handle all possible
types of gauge fixing procedure in a formalism which respects a full
symmetry between the ghosts and the antighosts.  This idea of adding to a
system new fields which count as a whole an effective number of degrees of
freedom equal to zero, with some possibilities of transferring the degrees
of freedom from one sector to the other could have further applications:
our general formula could be useful to couple purely topological theories
to ordinary gauge systems, with the purpose of of triggering phase
transitions by spontaneous breaking of either the ghost number conservation
or the topological BRST symmetry.



\end{document}